\documentclass[%
 preprint,
 amsmath,amssymb,
 aps,
]{revtex4-1}
\usepackage{xcolor}
\usepackage{graphicx}
\usepackage{dcolumn}
\usepackage{bm}
\usepackage{hyperref}
\usepackage{color}
\usepackage[T1]{fontenc}
\usepackage[latin9]{inputenc}

\begin{document}

\preprint{APS/123-QED}

\title{Thermal conduction in one dimensional $\Phi^{4}$ chains with colliding particles}

\author{Sankhadeep Bhattacharyya}
 \affiliation{Department of Mechanical Engineering, Indian Institute of Technology Kharagpur, West Bengal, India - 721302.}
\author{Puneet Kumar Patra}%
 \email{puneet.patra@civil.iitkgp.ac.in}
\affiliation{%
Department of Civil Engineering and Center for Theoretical Studies, Indian Institute of Technology Kharagpur, West Bengal, India - 721302}%

\date{\today}

\begin{abstract}
Studying thermal conduction in low dimensional systems, such as $\Phi^4$ chains, helps us in understanding the microscopic origins of Fourier's law. This work relaxes the assumption of point particles prevalent in the study of thermal transport characteristics in $\Phi^4$ chains. The particles of the modified chain, henceforth termed as the $\Phi^{4C}$ chain, can collide with each other. Collisions have been modelled by adding a short-ranged soft-sphere potential to the Hamiltonian of the $\Phi^4$ chain. The inclusion of soft-sphere potential drastically alters the thermal transport characteristics while still satisfying the Fourier's law: (i) at low temperatures, the temperature profile has negligible boundary jumps in $\Phi^{4C}$ chains vis-\'a-vis $\Phi^4$ chains, (ii) thermal conductivity of $\Phi^{4C}$ chains is significantly smaller than $\Phi^4$ chains at low temperatures, (iii) at high temperatures, $\Phi^{4C}$ chains have a higher thermal conductivity than $\Phi^4$ chains, and (iv) unlike $\Phi^4$ chains, where thermal conductivity keeps decreasing upon increasing temperature, $\Phi^{4C}$ chains show a unique behavior not observed in other momentum non-conserving chains -- thermal conductivity abruptly decreases first and then increases beyond an inversion temperature. Splitting the total heat current into the contributions of the harmonic and anharmonic inter-particle forces, reveals that the harmonic contributions decrease with increase in temperature. On the contrary, anharmonic contributions increase with rising temperature, and beyond the inversion temperature they overtake the harmonic contributions.  Exploring the dynamics of the two chains in Fourier space helps in identifying that the energy of the lowest modes redistribute to other modes of vibration much faster in $\Phi^{4C}$ chains than in $\Phi^4$ chains due to collisions. The quicker redistribution of the energy to higher modes is the reason behind smaller thermal conductivity in $\Phi^{4C}$ chains at low temperatures. The proposed $\Phi^{4C}$ chains have the features of both momentum conserving as well as momentum non-conserving systems, and may become an important tool to study thermal transport in real-life systems. 
\end{abstract}

\maketitle


\section{\label{sec:level1}Introduction}



Recent technological advances have enabled researchers to engineer low dimensional systems, such as quantum dots, electron gas, carbon nanotubes (CNTs), nanowires, graphene, etc., where the motion of particles is severely restricted in one or more dimensions. These low dimensional systems are associated with interesting thermal transport characteristics, for example, one observes length and temperature dependent thermal conductivity in CNTs \cite{fujii2005measuring} and a divergent thermal conductivity in graphene \cite{balandin2008superior}, which stand in stark contrast (and hence, anomalous) to the Fourier's law of thermal conduction, where $\kappa$, the thermal conductivity, is independent of length:
\begin{equation}
J = -\kappa{\nabla T}.
\end{equation}
Here, $J$ is the heat flux, and $T$, the local temperature.
In order to explain the anomalous behavior, researchers have probed into the thermal transport characteristics of idealized one-dimensional chains. Over the years, several such chains have been proposed. These may be classified into three different categories based on their thermal transport characteristics -- (i) chains with ballistic thermal conduction as seen in a chain of harmonically coupled oscillators \cite{lepri2003thermal}, (ii) chains displaying anomalous thermal conduction as is observed in a Fermi-Pasta-Ulam (FPU) chain \cite{lepri1997heat}, and (iii) chains where thermal transport characteristics obey Fourier's law as is observed in a $\Phi^4$ chain \cite{chen1996breather} and a Frenkel-Kontorova (FK) model \cite{hu2005heat}. In general, the Hamiltonian of one-dimensional chains may be written as:
\begin{equation}
    H = \sum\limits_{i=1}^{N} \left[\frac{p_i^2}{2m} + V_H(\Delta x_{i-1,i}) + V_A(\Delta x_{i-1,i}) + U(x_i) \right],
    \label{eq:gen-Hamiltonian}
\end{equation}
where $p_i$ is the momentum of the $i^{th}$ particle having a mass of $m$, $V_H$ is the harmonic part of the potential that depends on the distance between the two nearby particles $\Delta x_{i-1,i}$, $V_A$ is the anharmonic part of the potential that also depends on $\Delta x_{i-1,i}$ and $U(x_i)$ is the anharmonic tethering part of potential. Depending upon the choice of $V_H(\ldots), V_A (\ldots)$ and $U(\ldots)$, one ends up with different chains. For example, FPU chain is obtained when $V_A(\Delta x_{i-1,i}) = \dfrac{1}{4}c_1 \Delta x_{i-1,i}^4$ and $U(x_i) = 0$, while the $\Phi^4$ chain is obtained when $V_A = 0$ and $U(x_i) = \dfrac{1}{4}c x_i^4$, and with $V_A(\ldots) = 0, U( \ldots) = f(\cos(x_i))$ one obtains FK model. 

So why do the chains behave differently? The question has baffled researchers for several years. An important breakthrough was made by Casati et al \cite{casati1984one} through their ding-a-ling model, where every alternate particle was attached to its initial position with harmonic spring and the remaining ones were free. It was discovered that a key ingredient for normal thermal conductivity is chaos. However it was later found that although chaos is necessary, it is not sufficient for ensuring that the Fourier's law is obeyed. For example, the FPU-$\beta$ chain {\cite{lepri1997heat}, which is chaotic, has a thermal conductivity that diverges with system size, $L$, in a power law manner -- $\kappa \propto L^{0.3}$.

Looking at the specific form of Hamiltonian for the different chains, researchers observed a major difference between the chains -- while the FPU chain is momentum preserving, the other ones ($\Phi^4$, FK and ding-a-ling) are not -- and tried to link the momentum preserving characteristics with anomalous thermal transport behavior. Researchers initially attributed the anomalous thermal transport behavior in the momentum preserving chains to the slow diffusion of energy carried by low-frequency and long-wavelength modes \cite{prosen2000momentum}. These modes act as nearly undamped energy transport channels, and cause long-time correlations within the system. The presence of tethering potential in momentum non-conserving systems disrupts the energy transport of the long-wavelength modes, causing normal thermal transport behavior. However, recent research indicates that not all momentum conserving systems display divergent thermal conductivity \cite{savin2014thermal,giardina2000finite,gendelman2000normal,lee2010momentum,giardiana2005momentum}. For example, Wang et. al. \textcolor{black}{\cite{wang2013validity}} have shown through non-equilibrium molecular dynamics that their one-dimensional chain with asymmetric interparticle interactions has a convergent thermal conductivity in thermodynamic limit. Further, it was found that anomalous thermal conductivity can be obtained in systems without momentum conservation \textcolor{black}{\cite{prosen2000momentum}}. Interestingly, the thermal transport characteristics of the one-dimensional chains are dependent on the nature of coupling. For example, it has been observed that a $\Phi^4$ chain under weakly non-linear tethering potential displays ballistic thermal conduction \cite{xiong2017crossover}. Thus, it is evident that additional hidden attributes are at play in determining if a one-dimensional chain exhibits anomalous thermal transport.


The polynomial non-linearity in the Hamiltonian is the result of Taylor series truncation of the complete interaction potential \cite{gendelman2016heat}. While this truncation is a good approximation at low temperatures, it is not very realistic at high temperatures where two particles may come very close or go far from each other. Even at low temperatures the potential fails to account for the finite probability of two particles coming very close to each other. The point mass assumption coupled with low-order non-linear potential allows two particles to cross each other so that they never ``collide''. The situation can be made more realistic by incorporating a high-order non-linear potential in the total Hamiltonian. In this manuscript we consider a $\Phi^4$ chain and relax the assumption of point particles -- two particles are prevented from crossing each other and collide upon coming closer than a threshold distance. This is achieved by modifying the Hamiltonian of $\Phi^4$ chain to include a high-order soft sphere repulsive potential. Specifically, we seek the solutions to the following questions -- (i) what is the effect of collisions on thermal conductivity and heat flux, (ii) do collisions have any bearing on boundary effects in temperature profile, and (iii) how do collisions alter the diffusion of energy. Our results indicate that collisions significantly reduce thermal conductivity and heat flux, and remove boundary effects present in the temperature profile of non-colliding chains. The reason behind significant reduction in heat flux and thermal conductivity may be understood by looking at the dynamics in the Fourier space. Results indicate that energy diffuses from the lowest modes to higher modes faster when collisions are incorporated within the system. 

This manuscript is organized as follows: the next section details the modification made to the $\Phi^4$ Hamiltonian to obtain the $\Phi^{4C}$ chain, subsequently we highlight the simulation methodology adopted in the present study. Lastly, we present the results and conclusions of our study.


\section{The $\Phi^4$ and $\Phi^{4C}$ Chains}
A typical $\Phi^4$ chain comprises of $N$ particles, each of mass $m$, arranged on a one-dimensional line and separated by a distance $l_{eq}$. Each particle is connected with its nearest neighbour by means of a harmonic spring and to its initial equilibrium position by a quartic tethering spring. As a result, the harmonic and tethering potentials take the form: $V_H (\Delta x_{i-1,i}) = \dfrac{1}{2} k(x_{i-1}-x_i-l_{eq})^2$ and $U(x_i) = \frac{1}{4}c(x_i - x_{i,0})^4$, respectively. Here, $x_i$ and $x_{i,0}$ are the instantaneous and equilibrium positions of the $i^{th}$ particle, respectively. The boundary conditions may be taken as fixed, wherein a fixed particle of similar characteristics is placed at the either ends, or periodic, wherein the $N^{th}$ particle is connected with the first particle. For a $\Phi^4$ chain with fixed boundaries, and $m = k = l_{eq} = 1.0$, the Hamiltonian becomes:
\begin{equation}
\begin{array}{rcl}
H_{\Phi^4} & = & \sum\limits_{i=1}^{N} \left[\frac{p_i^2}{2} \right] + \sum\limits_{i=1}^{N-1} \left[\frac{1}{2} \left( x_{i} -  x_{i-1} - 1.0 \right)^2 \right] \\
& & + \sum\limits_{i=1}^{N} \left[ \frac{c}{4} \left( x_i - x_{i,0} \right) ^ 4 \right]
\label{eq:3}
\end{array}
\end{equation}
$\Phi^4$ chains with large anharmonicity ($c=1.0$) have been studied extensively by several researchers \cite{hu2000heat, aoki2000bulk, PhysRevE.93.033308} using deterministic thermostats and have been extended to more than one dimensions \cite{aoki2000non}. The momentum conservation breaks down in presence of tethering potential with momentum dissipating exponentially in time \cite{hu2000heat}. In this large $c$ limit, the heat flux, $J$, in the traditional $\Phi^4$ chain has been found to be inversely proportional to $N: J \sim 1/N$, suggesting that the chain exhibits normal thermal transport characteristics. Extensive numerical simulations, with $k = c = 1$, suggest that the thermal conductivity, $\kappa$, depends on the temperature, $T$, through the relation: $\kappa \approx 1.724/T^{1.382}$ \cite{aoki2000non}. 

\begin{figure}[h]
    \centering
    \includegraphics[scale=0.95]{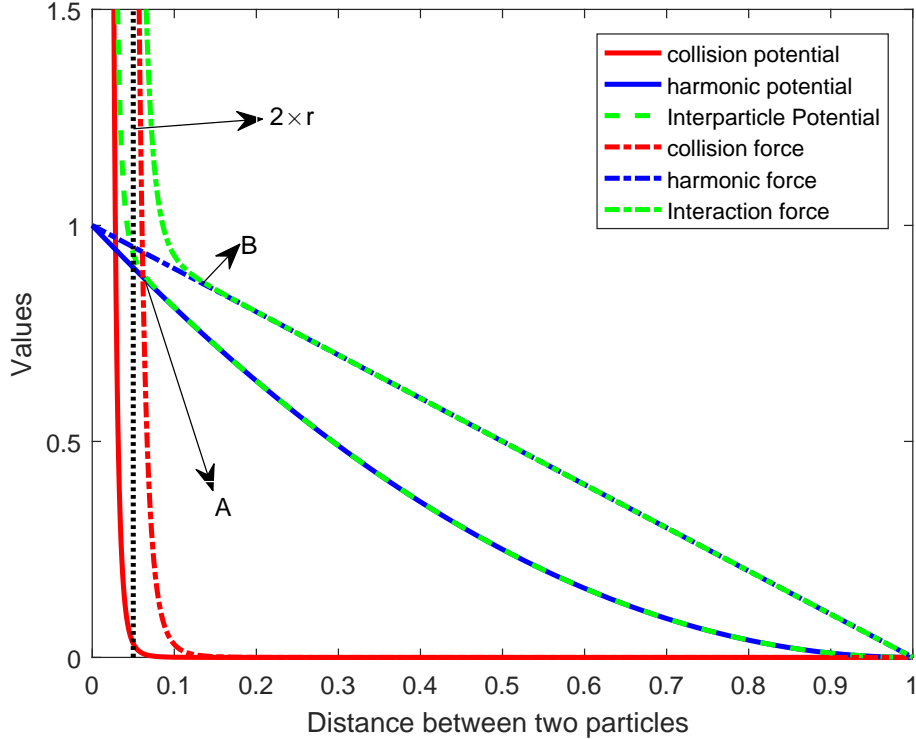}
    \caption{Different potential and force functions for a particle located at origin assuming a neighbour
particle at +1 units. Dotted line (parallel to $y$ axis) shows $2r$ beyond which, the soft sphere
potential becomes insignificant. A:(0.065,0.88) is where net potential deviates from harmonic,
B:(0.135,0.868) is where net force deviates from the harmonic force.}
    \label{fig:fig1}
\end{figure}
The traditional $\Phi^4$ model treats all particles as point masses and does not prevent two nearby particles from crossing each other. This situation is less realistic considering the fact that two atoms or molecules experience large repulsive forces upon coming very close to each other. In the present manuscript, we propose the $\Phi^{4C}$ chain whose Hamiltonian consists of an extra anharmonic term to take into account soft-sphere like collisions: $V_A( \ldots ) = -a\frac{1}{(x_i - x_{i-1})^6}$. The $\Phi^{4C}$ Hamiltonian reads:
\begin{equation}
\begin{array}{rcl}
H_{\Phi^{4C}} & = & \sum\limits_{i=1}^{N} \left[\frac{p_i^2}{2m} \right]+ \sum\limits_{i=1}^{N-1} \left[\frac{1}{2} \left( x_{i} -  x_{i-1} - l_{eq} \right)^2 \right] \\
& & - \sum\limits_{i=1}^{N-1} \left[ \frac {a}{(x_i - x_{i-1})^6} \right] \\
& & + \sum\limits_{i=1}^{N} \left[ \frac{c}{4} \left( x_i - x_{i,0} \right) ^ 4 \right]
\end{array}
\label{eq:4}
\end{equation}
$V_A( \ldots )$ ensures that two particles do not cross each other, and instead collide over a brief time period. 

For the remainder of this manuscript, $c=0.1$ and $a = 5 \times 10^{-10}$. A small value of $c$ ensures that the anharmonic tethering energy is only a fraction of the harmonic spring energy for the majority of simulation time. Our choice of the constant $a$ is governed by: (i) the effective radius of the particles, taken as $r = 0.025$, so that when the distance between two particles is smaller than $2r$, a very large repulsive force is experienced by the particles, and (ii) $V_A \to 0$ when the distance between the two particles is greater than $2r$ so that there is small deviation of the Hamiltonian from the traditional $\Phi^4$ Hamiltonian at these distances. Corresponding to the chosen values of $r$ and $a$, figure \ref{fig:fig1} shows the relative contribution of the soft-sphere potential (force) vis-\'a-vis the harmonic potential (force). As is evident from the figure, beyond a distance of 0.050, the interparticle potential and force are dominated by the harmonic potential and the corresponding force, respectively.

We now describe the methodology adopted in the present study.

\section{Simulation Methodology}
The particles of both $\Phi^4$ and $\Phi^{4C}$ chains are initialized such that their $i^{th}$ particle is located at $x_{i,0} = i-1$. Fixed-fixed boundary conditions are imposed on the chains through two fictitious stationary particles located at $x=-1$ and $x=N$. These boundary particles interact with the remaining particles by means of $V_H(\ldots) + V_A (\ldots)$ as described in the previous section. The initial velocity of each particle is chosen randomly from a uniform distribution between -0.5 to 0.5. A thermal gradient is effected on the system by keeping the first and the last particle of the chain in contact with heat reservoirs maintained at temperatures $T_H$ and $T_C$, respectively, where $T_H > T_C$. A graphical representation of the chains is shown in figure \ref{fig:fig2}. 
\begin{figure}
    \center
    \includegraphics[scale=0.5]{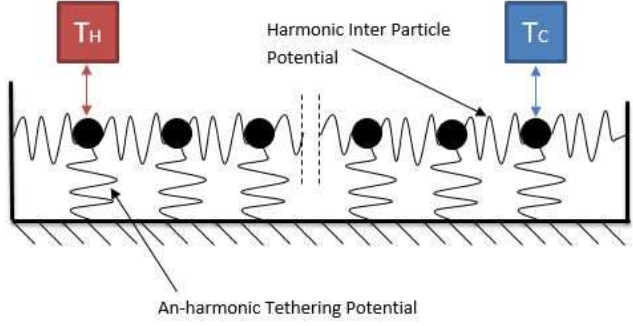}
    \caption{Pictorial depiction of the chains with the fixed-fixed boundary condition. Here, $T_H (T_C)$ denotes the hot (cold) thermal reservoir.}
    \label{fig:fig2}
\end{figure}

Due to the simplicity and wide adoption in scientific studies, from amongst the different deterministic thermostat algorithms \cite{nose1984unified,martyna1992nose,patra2014deterministic,patra2015deterministic,braga2005configurational} for controlling temperature of the reservoirs, we choose two Nos\'e-Hoover thermostats \cite{hoover1985canonical} -- one for $T_H$ and another for $T_C$. For both the thermostats, the thermostat mass was chosen as unity. The intermediate particles present between the first and the last particle are governed by standard Hamiltonian evolution. Thus, the resulting equations of motion are: 
\begin{equation}
    \begin{array}{rcl}
         \dot{q_i} & = & \frac{\partial{H}}{\partial{p_i}} \\
         \dot{p_i} & = & -\frac{\partial{H}}{\partial{q_i}} - \delta_i^H\zeta_H p_1 - \delta_i^C\zeta_C p_N\\
         \dot{\zeta_H} & = & \frac{p_1^2}{T_H} - 1\\
         \dot{\zeta_C} & = & \frac{p_N^2}{T_C} - 1\\
         \text{where :}\\
         \delta_i^H & = & 1 \{\text{if } i = 1 \}\\
         \delta_i^C & = & 1\{ \text{if } i = N \}\\ 
    \end{array}
    \label{eq:eq5}
\end{equation}
These equations of motion are solved using the $4^{th}$ order Runge-Kutta method. For the $\Phi^4$ chain, the incremental time-step, $\Delta t$, is chosen as 0.0005, and the system is allowed to evolve for 1 billion time steps. The first 250 million time steps are for ensuring that steady-state sets in within the chain while the last 750 million time steps are the actual runs from which all time averages are computed. For the $\Phi^{4C}$ chains, owing to the differential equations being stiff, the incremental time-step is halved to $\Delta t = 0.00025$. The $\Phi^{4C}$ equations are solved for 2 billion time steps, of which the first 500 million time steps bring the chain to steady-state conditions and the last 1.5 billion time steps are used for computing time averages. 

Both $\Phi^4$ and $\Phi^{4C}$ chains with $N= 16, 32, 64, 128, 256, 384$ and $512$ particles have been considered to understand the length scaling behavior of thermal conductivity. Keeping $T_H = 1.10T_M$ and $T_C = 0.90T_M$, we consider four values of $T_M=2.0,1.0,0.50$ and 0.10 for each case to identify the temperature scaling behavior of thermal conductivity. For reasons that will be apparent later, additional simulations have been performed for $\Phi^{4C}$ chains with $N=512$ and $T_M$ varying from 0.1 to 1.0 in increment of 0.05.

\subsubsection{Temperature and Thermal Conductivity Computation:}
Under the assumption that local thermodynamic equilibrium conditions \cite{patra2014deterministic} prevail within a chain, it is possible to define the different thermodynamic properties for every particle of the chain, including temperature and energy current. Temperature of a particle may be defined in multiple ways -- kinetic, configurational, Rugh's, etc. -- all of which are same under local thermodynamic equilibrium conditions \cite{patra2017nonequilibrium}. Consequently, we choose the simplest way of defining temperature -- the kinetic temperature -- as the temperature of a particle:
\begin{equation}
    k_B T_i = \langle mv_i^2 \rangle,
\end{equation}
Here, $T_i$ is the kinetic temperature of the $i^{th}$ particle. For the remainder of this study, the Boltzmann constant $k_B$ is set at unity. $\langle \ldots \rangle$ denotes the long time averaged value. Interested readers are referred to the review papers by Powles et. al \cite{powles2005temperatures} and Casas-V\'azquez and Jou \cite{casas2003temperature} for other ways of defining temperature.


The instantaneous local heat current at the $i^{th}$ site may be obtained by taking the time derivative of the local energy density, $\epsilon_i$, associated with the $i^{th}$ particle \cite{dhar2008heat}, and can be written as:
   \begin{equation}
     \begin{array}{rcl}
     \dot{\epsilon_i}& = &\frac{\partial\epsilon_i}{\partial t} +\left[ j_{i-1,i} -j_{i,i+1} \right].
     \end{array}
\label{eq:eq7}
 \end{equation}
Here, $j_{i,j}$ is the energy current flowing from the $i^{th}$ to the $j^{th}$ particle, and is given by: $j_{i,j} = \dfrac{1}{2} \left[ f_{i,j}(v_i+v_j) \right]$. Note that $f_{i,j}$ is the sum of harmonic and anharmonic forces acting on the $j^{th}$ particle due to the $i^{th}$ particle. At steady-state, where the time averaged quantities, $\langle \dot{\epsilon_i} \rangle= \langle \frac{\partial\epsilon_i}{\partial t} \rangle = 0$, equation (\ref{eq:eq7}) simplifies to $\langle j_{i-1,i} \rangle = \langle j_{i,i+1} \rangle$. Further, at steady state since  $\langle d V (\Delta x_{i-1,i})/ dt \rangle = 0$, we get \cite{dhar2008heat}:
 \begin{equation}
     \begin{array}{rcl}
     \langle j_{i-1,i} \rangle& = \langle \frac{1}{2}(v_i+v_{i-1})f_{i-1,i} \rangle = \langle v_if_{i-1,i} \rangle.
     \end{array}
 \end{equation}
The time averaged value of heat flux, $\langle J \rangle$, may now be computed as:
\begin{equation}
\langle J \rangle = \left\langle\dfrac{\sum\limits_{i=1}^{N} j_{i,i-1}}{N}\right\rangle,
\end{equation}
from which the thermal conductivity, $\kappa$, emerges as:
\begin{equation}
    \begin{array}{rcl}
    \kappa & = & \dfrac{\langle J \rangle N}{\Delta T}.\\
    \end{array}
    \label{eq:kappa}
\end{equation}
Here, $\Delta T / N$ is the temperature gradient with $\Delta T = T_H - T_C = 0.20 T_M$.

\subsubsection{\label{sec3.3}Monitoring of Modal Energy:}
In order to understand the energy transfer between modes, a separate set of simulations have been performed to monitor the energy of each mode in constant energy ensemble wherein, the equations of motion (\ref{eq:eq5}), simplify to: $\dot{q}_i = \dfrac{\partial H}{\partial p_i}, \dot{p}_i = \dfrac{-\partial H}{\partial q_i}$. Calculation of modal energy requires the knowledge of the modal parameters -- mode shapes, modal frequencies, modal displacements and modal velocities. 

Neglecting the anharmonic part of the potential, all modal parameters can be obtained by diagonalizing the mass-normalized Hessian matrix, $\mathbf{[M^{-1}K]}$. For a chain comprising $N$ particles, $\mathbf{[K]}$ is a symmetric matrix of dimension $N\times N$, whose elements are:
\begin{equation}
K_{i,j} = \dfrac{\partial^2 V_H}{\partial x_i \partial x_j} \to K_{i,i} = 2, K_{i,i+1} = K_{i-1,i} = -1.
\label{eq:5}
\end{equation}
All remaining terms are zero. Since, the mass matrix, $\mathbf{[M]}$, is an identity matrix for our case, the diagonalization of $\mathbf{[K]}$ provides the normal modal frequencies, $\omega_i^2, i \in [1,N]$, and the corresponding normal modes, $\vec{\xi}_i,i \in [1,N]$.


The instantaneous modal displacement, $\delta_i(t)$, and velocity, $\dot{\delta_i(t)}$, corresponding to the $i^{th}$ mode of vibration may be obtained by projecting the instantaneous displacements and velocities of all $N$ particles onto the $i^{th}$ eigenvector:
\begin{equation}
\begin{array}{rcl}
\delta_i(t) & = & \sum\limits_{j=1}^{N} (x_j(t) - x_j(0))\xi_{ij}, \\
\dot{\delta}_i(t) & = & \sum\limits_{j=1}^{N} \dot{x}_j(t) \xi_{ij}, \\
\end{array}
\label{eq:7}
\end{equation}
Thus, the instantaneous energy of the $i^{th}$ normal mode becomes:
\begin{equation}
E_i(t) = PE_i(t) + KE_i(t) = \dfrac{1}{2}\omega_i^2 \delta_i^2(t) + \dfrac{1}{2} \dot{\delta}_i^2(t),
\label{eq:6}
\end{equation}
where, $PE_i$ and $KE_i$ are the potential and kinetic energies of the $i^{th}$ mode, and equal $\dfrac{1}{2}\omega_i^2 \delta_i^2(t)$ and $\dfrac{1}{2} \dot{\delta}_i^2(t)$, respectively.

The following steps are used for continuously monitoring the modal energy:
\begin{enumerate}
\item At the beginning of the simulation, the matrix $\mathbf{[K]}$ is obtained using equation (\ref{eq:5}), and its mass-weighted form is diagonalized to obtain $\omega_i , i \in [1,N]$ and $\vec{\xi}_i, i \in [1,N]$.
The chains are initialized such that the initial energy is concentrated in the first mode. This is obtained by imparting a velocity to all the particles according to the first eigenvector, $\vec{\xi}_1$:
\begin{equation}
    \begin{array}{rcl}
         v_i&=&\alpha \xi_{1,i}.
    \end{array}
\end{equation}
Here $\alpha$ is a scaling constant taken as 10.

\item \label{loop:a} Equations of motion for the chains are solved in a time-incrementing loop:
\begin{enumerate}
\item  Update the position and velocity of all particles as per the equations of motion.
\item Compute $\delta_i$ and $\dot{\delta}_i$ using the expressions (\ref{eq:7}).
\item Evaluate modal energy of all modes using the expression (\ref{eq:6}).
\item Repeat \ref{loop:a} until the required time is reached. 
\end{enumerate}
\end{enumerate}
The equations of motion are solved for 200,000 time steps for $\Phi^4$ chains with $\Delta t = 0.0005$, while for $\Phi^{4C}$ chains, the equations of motion are solved for 400,000 time steps with $\Delta t = 0.00025$. Note that neglecting the anharmonic contributions makes  $\omega_i , i \in [1,N]$ and $\vec{\xi}_i, i \in [1,N]$ constant throughout the simulations, and hence, need to be evaluated only once.

\section{Results:}

\subsection{Verification}
In order to check the veracity of our simulations, we compare previously reported solutions of $\Phi^4$ chains with those obtained from our code. The test cases correspond to $k=c=1.0$ and $N=512$ particles. The simulation methodology has been kept the same as highlighted before including the boundary conditions. The thermostats were set at three different mean temperatures $T_M  = 1.0, 0.5$ and 0.1 with the higher temperature, $T_H$, being $10\%$ more and the lower temperature, $T_C$, being $10\%$ lesser than $T_M$. Theoretically, the conductivity follow $\kappa_{th} = 2.724/T_M^{1.382}$ \cite{aoki2000non}. The values obtained from the simulations have been compared to the theoretical values in table \ref{table1}, and as can be seen, there is a good agreement between $\kappa_{th}$ and $\kappa$. The difference between them at lower temperatures occurs due to the different boundary conditions used.

\begin{table}
 \caption{\label{table1} A $\Phi^4$ chain with $k=1$ and $c=1$ is subjected to thermal conduction as per the methodology highlighted before. Our simulation results for thermal conductivity ($\kappa$) are compared with those obtained by Aoki and Kusnezov \cite{aoki2000non}: $\kappa_{th} = 2.724/T_M^{1.382}$. The difference in the results arise because of different boundary conditions adopted for finding $\kappa_{th}$.}
\begin{center}
 \begin{tabular}{c c c c c} 
 \hline
 $T_M$ & $\langle J \rangle$  & $J\times N$ & $\kappa$ & $\kappa_{th}$  \\ \hline
 \hline
 1.0 & 0.0011 & 0.570 & 2.849 & 2.724 \\ 
 0.5 & 0.0013 & 0.647 & 6.470 & 7.100 \\
 0.1 & 0.0023 & 1.574 & 57.870 &65.646 \\ [1ex] 
 \hline
\end{tabular}
\end{center}
\end{table}

\subsection{Check for steady-state conditions}
In non-equilibrium settings, meaningful time averages can be taken only after a system reaches steady state conditions. As described before, we have assumed that the chains reach steady-state conditions after 250 million time steps. We now check if our assumption is valid. At steady state, the net heat current, $\langle J \rangle$, must equal local heat current flowing between any two adjacent particles i.e. $\langle j_{1,2} \rangle = \ldots = \langle j_{i-1,i} \rangle = \langle j_{i,i+1} \rangle = \ldots = \langle j_{N-1,N} \rangle$. A significant deviation from this condition indicates that the system has not yet reached a steady state. Resetting the parameter $c$ to 0.1 in $\Phi^4$ chains with $N=512$ particles, we compute the following time-averaged quantities for the 750 million actual simulation runs: $(\langle J \rangle, | \langle J \rangle - \min \langle j_{i-1,i} \rangle|, | \langle J \rangle - \max \langle j_{i-1,i} \rangle |)$. The results for $T_M = 1.0, 0.5$ and $0.1$ are: $(0.0221,4.71 \times 10^{-5},5.88 \times 10^{-5}), (0.0237,3.383\times 10^{-5},2.366 \times 10^{-5}),(0.009955,1.4206\times 10^{-5},7.7032\times 10^{-6})$, respectively. The small deviations are indicative of the fact that one can take time averages for the 750 million actual simulation meaningful runs.

\subsection{Temperature Profile}
\begin{figure}
    \centering
    \includegraphics[scale=0.5]{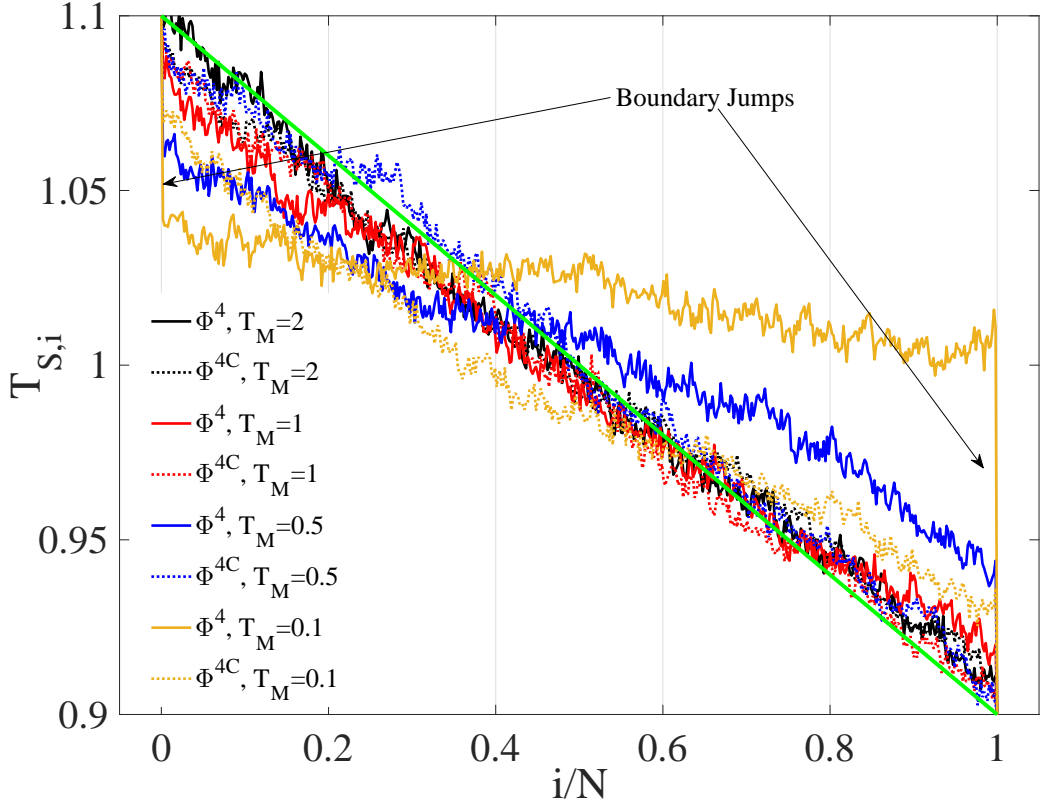}
    \caption{Scaled temperature profile of $\Phi^4$ and $\Phi^{4C}$ chains with N = 512 particles subjected to four different values of $T_M$. Scaling is done such that the scaled values of $T_H $ and $T_C$ are 1.1 and 0.9, respectively. Scaled temperature, $T_{S,i}$, of the remaining particles are interpolated. As can be seen from the figure, $\Phi^4$ particles exhibit larger boundary jumps in the temperature profile than the $\Phi^{4C}$ particles. In absence of these boundary jumps, the $\Phi^{4C}$ chain deviates marginally from a linear profile (green line) even at low $T_M$. On the other hand, in $\Phi^4$ chains, the presence of boundary jumps causes significant deviation from linear temperature profile at low $T_M$. }
    \label{fig:fig3}
\end{figure}

Assuming that local thermodynamic equilibrium conditions prevail within the chain under the prescribed temperature gradient, each particle of the chain has a well defined kinetic temperature. In order to make a uniform comparison of temperature profiles, the kinetic temperature of each particle, $T_i$, is normalized with $T_M$:
\begin{equation}
    T_{S,i} = \dfrac{T_i}{T_M}
\end{equation}
Figure \ref{fig:fig3} depicts the variation of $T_{S,i}$ for the $\Phi^4$ and $\Phi^{4C}$ chains with $N=512$ particles. Introducing soft-sphere collision potential to the $\Phi^4$ chain drastically alters the temperature profile. A typical $\Phi^4$ chain exhibits boundary jumps in temperature profile \cite{aoki2001fermi}, which become more pronounced with decreasing $T_M$. In comparison, due to the soft-sphere collision potential, the boundary jumps are negligible in the $\Phi^{4C}$ chains. The difference between the two chains become markedly noticeable at lower values of $T_M$. We point to the readers that the exact reason for such boundary jumps is yet to be found. 

The linearity of temperature profile is a key signature of normal thermal transport characteristics \cite{lepri2016thermal}. It has been previously found that the temperature profile (away from the boundary jumps) in $\Phi^4$ chains varies linearly \cite{aoki2000non}. We now compute the deviation from linearity, $d_L$, for both the chains through: 
\begin{equation}
    d_L = \sum\limits_{i=1}^{N} \sqrt{\left( T_{S,i} - Y_i \right)^2},
\end{equation}
where, $Y_i$ denotes the ordinate of the straight line corresponding to the $i^{th}$ particle. The results of $d_L$ are shown in table \ref{tab:table2} and confirm that, at low $T_M$, the temperature profile in $\Phi^{4C}$ chains is closer to being a straight line than that in $\Phi^4$ chains. Interestingly enough, while the deviation from linearity keeps decreasing with increasing $T_M$ for $\Phi^4$ chains, such a trend remains absent in the $\Phi^{4C}$ chains. The reduction in boundary jumps and deviation from linearity suggests that the $\Phi^{4C}$ chain allows for quicker thermalization and mimics macroscopic behavior better than the standard $\Phi^4$ chain at lower temperatures.
\textcolor{blue}{
\begin{table}
 \caption{A comparison of deviation from linearity, $d_L$, for $N=512$ particles and four values of $T_M$. As is evident, the deviation from linearity is lesser for $\Phi^{4C}$ chains at low $T_M$. }
\begin{center}
 \begin{tabular}{c c c} 
 \hline
 $T_M$&$\Phi^4$ & $\Phi^{4C}$\\
 \hline\hline
 2.0&0.1529&0.1951\\
 \hline
 1.0&0.279&0.1923\\
 \hline
 0.5&0.603&0.1450\\
\hline
 0.1&1.175&0.5060 \\ [1ex] 
 \hline
\end{tabular}
\end{center}
\label{tab:table2}
\end{table}
}

\subsection{Thermal conductivity}
\begin{figure}
    \centering
    \includegraphics[scale=0.5]{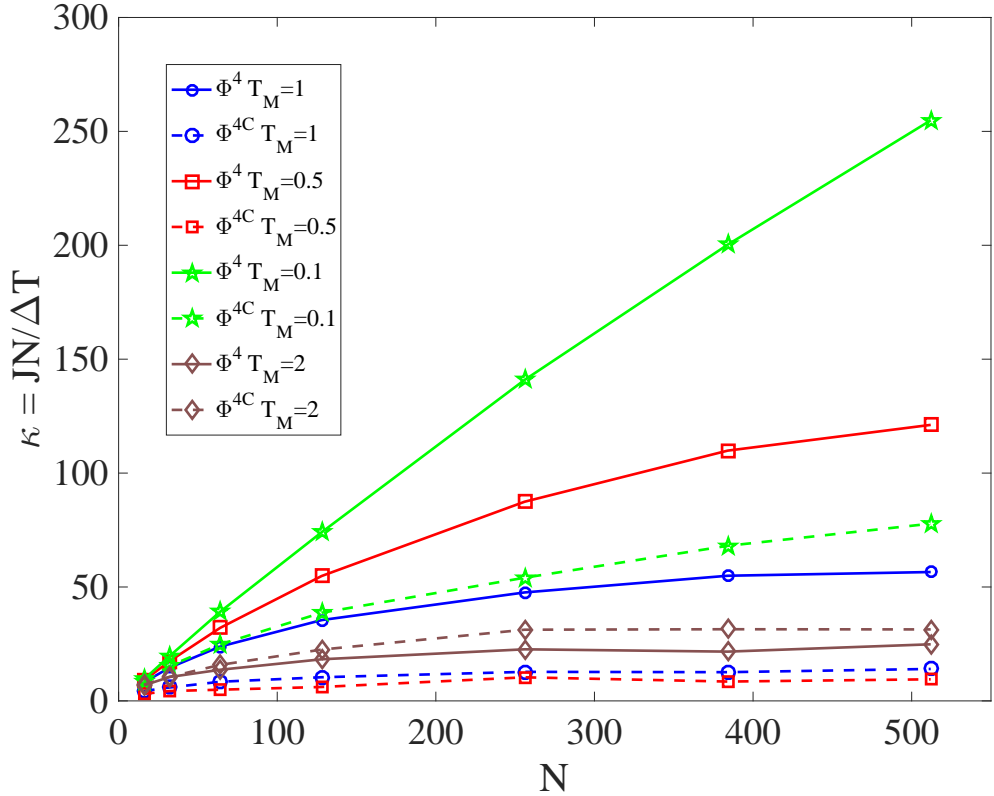}
    \caption{Variation of thermal conductivity, $\kappa$, with $N$ and $T_M$ for $\Phi^4$ and $\Phi^{4C}$ chains. While in the $\Phi^4$ chain, $\kappa$ increases with decreasing temperature owing to harmonic effects being predominant, the behavior of $\Phi^{4C}$ chain is not so straight forward. It first decreases with temperature and then suddenly increases. We attribute this behavior to the increased collisions at higher temperature. Interestingly enough, while for small $T_M$, $\kappa$ of $\Phi^{4C}$ chains is significantly smaller than that of $\Phi^4$ chains, such is not true for $T_M = 2.0$, where, $\kappa$ of $\Phi^{4C}$ chains is larger than $\Phi^4$ chains. }
    \label{fig:fig4}
\end{figure}

Thermal conductivity, $\kappa$, of both $\Phi^4$ and $\Phi^{4C}$ chains (shown in solid and dashed lines, respectively) are plotted in figure \ref{fig:fig4} for different values of $T_M$ and $N$. In $\Phi^4$ chains, $\kappa$ decreases with increasing $T_M$ and conforms with previously reported results \cite{PhysRevE.61.3828}. The reason may be attributed to the increased contribution of the anharmonic part of the potential because of the enhanced vibrations of the individual particles at higher temperatures. Consequently, the different modes of vibration interact with each other, and energy is transferred from the lower modes to the higher modes. In contrast, at low $T_M$ where harmonic effects dominate, the temporal evolution of energy in lower modes occurs nearly unimpeded, resulting in near-ballistic thermal conduction (see the solid green line of figure \ref{fig:fig4} ). 

Like $\Phi^4$ chains, $\Phi^{4C}$ chains satisfy Fourier's law as is evidenced by the plateauing of the graphs with increasing $N$ for a specific value of $T_M$. Further, it is apparent from figure \ref{fig:fig4} that $\kappa$  in $\Phi^{4C}$ chains follows a unique trend with temperatures which is not seen in any other one-dimensional momentum non-conserving chains -- as $T_M$ increases from 0.1 to 0.5, $\kappa$ reduces quickly, however, upon increasing $T_M$ further, $\kappa$ starts to rise. In order to better understand this phenomenon, the variation of $\kappa$ with $T_M$ is plotted in figure \ref{fig:fig5} for $N=512$ particles as $T_M$ is increased from 0.1 to 1.0 in increments of 0.05. It follows from figure \ref{fig:fig5} that beyond the inversion temperature of $T_M \approx 0.55$, $\kappa$ increases. 
\begin{figure}
    \centering
    \includegraphics[scale = 0.5]{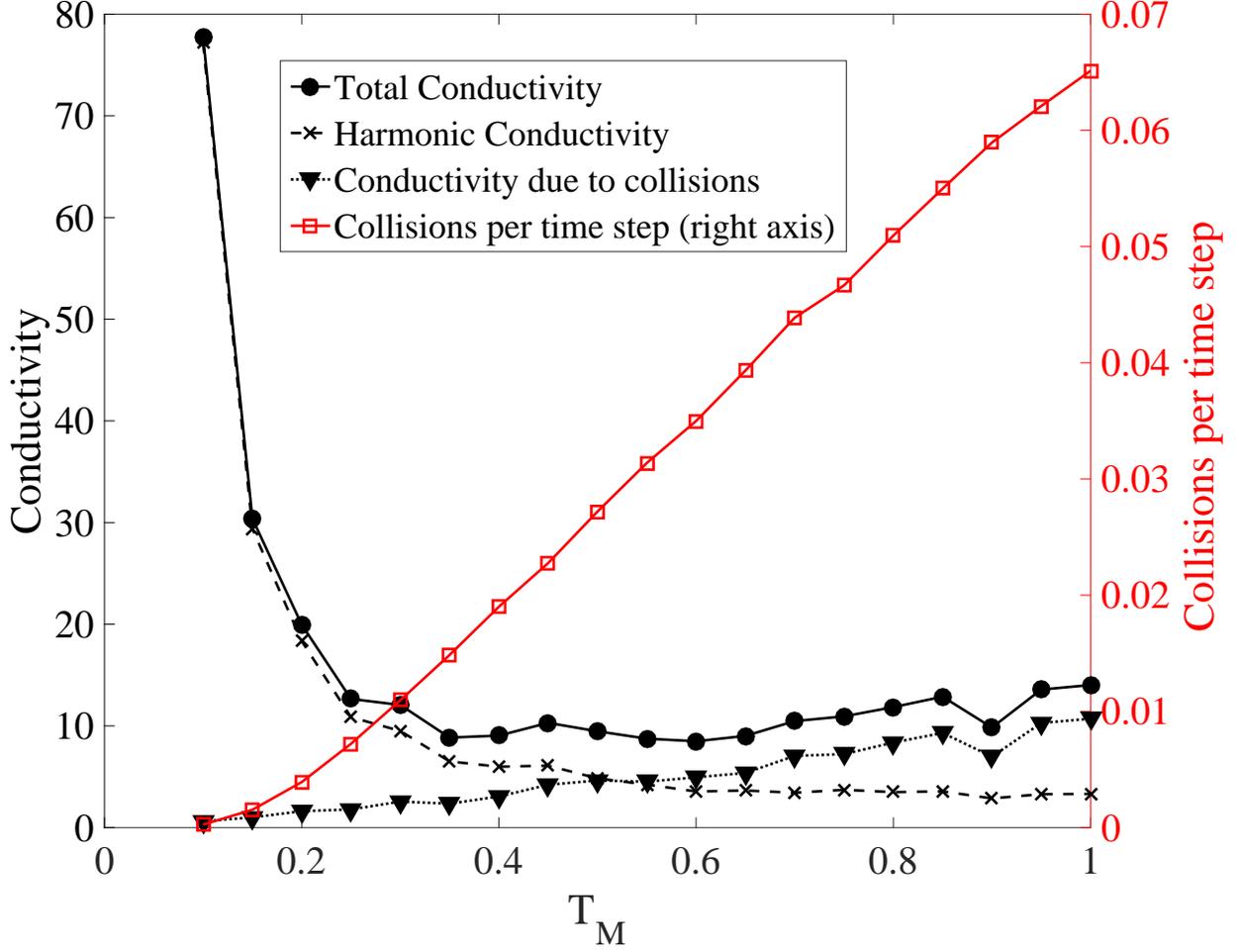}
    \caption{Variation of total thermal conductivity ($\kappa$), harmonic and anharmonic contributions to $\kappa$ for a $\Phi^{4C}$ chain comprising $N=512$ particles as $T_M$ is increased from 0.1 to 1.0. With increasing $T_M$, the particles tend to vibrate more and collide more frequently. Consequently, the contribution of harmonic part of the potential towards $\kappa$ decreases and the anharmonic contribution of soft-sphere potential starts to increase. Upon crossing the inversion temperature of $T_M \approx 0.55$, the anharmonic contribution overshadows the harmonic contribution.}
    \label{fig:fig5}
\end{figure}

The contributions to $\kappa$ may be split into harmonic ($J_H$) and anharmonic ($J_A$) heat currents, which can further be related to the harmonic ($f_{H,i,i-1}$) and anharmonic ($f_{A,i,i-1}$) inter-particle forces:
\begin{equation}
\begin{array}{rcl}
    \kappa & = & \dfrac{\langle J_H \rangle N}{\Delta T} + \dfrac{\langle J_A \rangle N}{\Delta T} \\
           & = & \dfrac{\langle \sum j_{H,i,i-1} \rangle}{\Delta T} + \dfrac{\langle \sum j_{A,i,i-1} \rangle}{\Delta T} \\
           & = & \dfrac{\langle \sum v_i f_{H,i,i-1} \rangle}{\Delta T} + \dfrac{\langle \sum v_i f_{A,i,i-1} \rangle}{\Delta T} \\
\end{array}
\end{equation}
The anharmonic part of the heat current, $J_A$ has a role to play when two particles ``collide''. It must be noted that since two particles of a $\Phi^{4C}$ chain never actually undergo head-on collisions in presence of soft-sphere potential, we identify a collision event from the trajectory of two particles -- the collision count is incremented by one whenever the relative velocity of the two particles gets reversed as they come within the effective radius, $r$. At higher $T_M$, collisions are more frequent, as can be seen from the secondary $y$ axis of figure \ref{fig:fig5} which plots the number of such collisions occurring per time step. The unique trend of $\kappa$ in $\Phi^{4C}$ chains occurs because $J_A$ increases with increasing frequency of collisions, so much so, that beyond $T_M \approx 0.55$, it overtakes $J_H$. 


\subsection{Normal Modes}
\begin{figure}
    \centering
    \includegraphics[scale = 0.5]{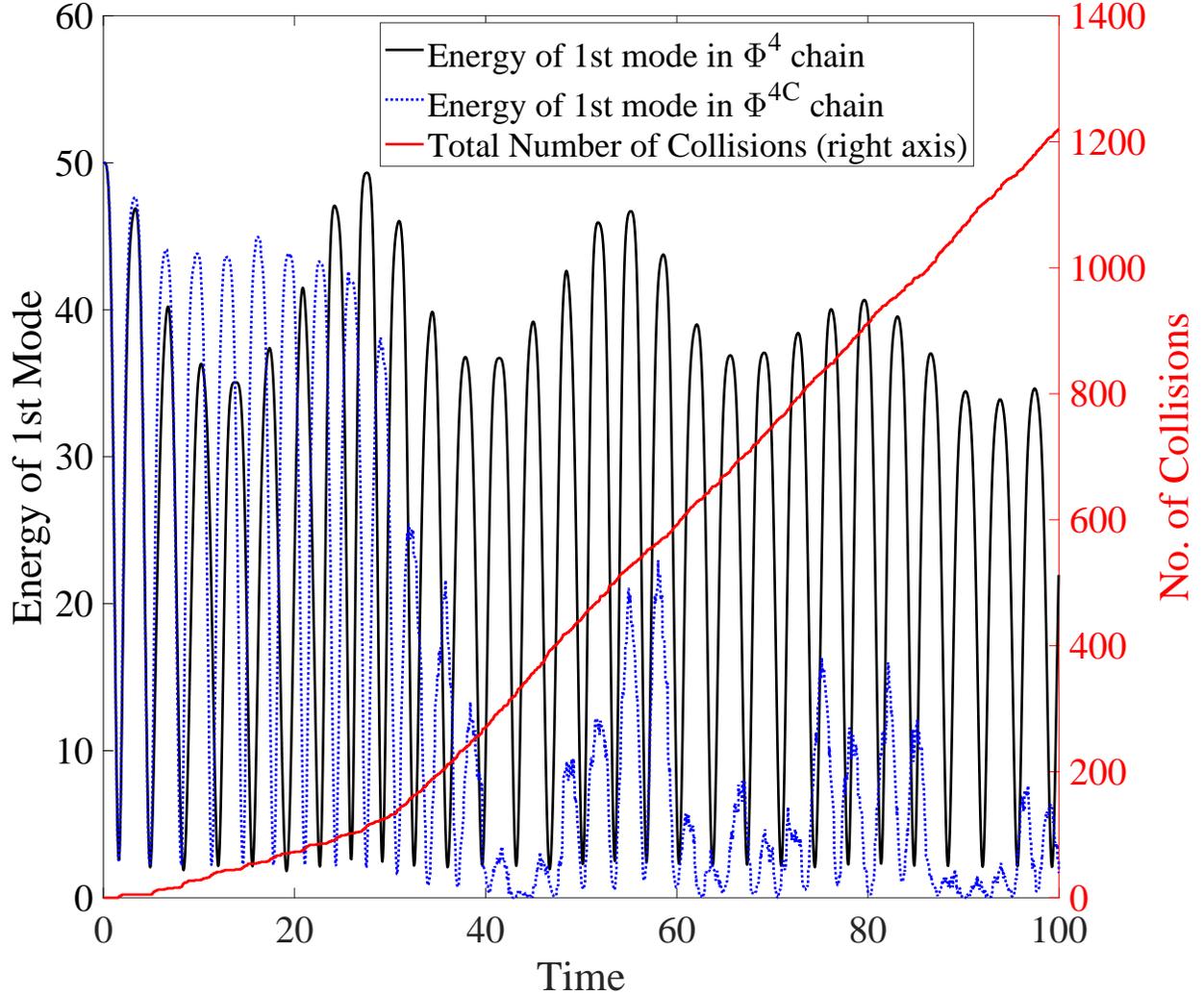}
    \caption{Temporal evolution of the first normal modal energy for $\Phi^4$ and $\Phi^{4C}$ chains comprising of $N=16$ particles. The methodology adopted for these simulations is discussed in section \ref{sec3.3}. The system is initialized so that all energy is concentrated in the first mode of vibration. As time progresses, it is evident that the the energy of first mode is distributed to other modes. However, the speed of redistribution is much faster in $\Phi^{4C}$ chains than in the $\Phi^4$ chains because of collisions. Due to the quicker redistribution of modal energy in the $\Phi^{4C}$ chains, the thermal conductivity in $\Phi^{4C}$ chains is much smaller initially.}
    \label{fig:fig6}
\end{figure}
It is now well known that high thermal conductivity in one-dimensional chains occurs because of the slow diffusion of energy carried by the low-frequency and long-wavelength modes. In presence of anharmonicity (due to tethering potential in $\Phi^4$ chains and tethering + soft-sphere collision potentials in $\Phi^{4C}$ chains), the lowest modes interact with the higher modes, causing energy transfer from the lowest modes. Thus, in both $\Phi^4$ and $\Phi^{4C}$ chains, one observes finite thermal conductivity. However, the rate of energy transfer from the lowest modes is different for the two chains. From figure \ref{fig:fig4}, it is evident that at low $T_M$, thermal conductivity in $\Phi^{4C}$ chains is significantly smaller than in $\Phi^4$ chains. This suggests that the rate of energy transfer from the lowest modes is significantly faster in $\Phi^{4C}$ chains. 

In order to justify our statement, we now look into the temporal evolution of normal modal energies of both the chains using the methodology highlighted in section \ref{sec3.3}. Figure \ref{fig:fig6} plots the temporal evolution of the first normal modal energy with $N=16$ particles. It can be observed from the figure that the time averaged energy of the first mode decreases faster in $\Phi^{4C}$ chains than in $\Phi^4$ chains. The strong anharmonicity occurring during collisions causes the energy of the first mode to quickly redistribute itself to higher modes, significantly reducing thermal conductivity. 

The relationship between thermal conduction and energy redistribution amongst the different modes gets somewhat obscure at high $T_M$. This is evident from the brown curves of figure \ref{fig:fig4}, where it can be seen that at $T_M = 2.0$, $\Phi^{4C}$ chains have a larger thermal conductivity than $\Phi^4$ chains. At higher temperatures, because of increased frequency of collisions, the anharmonic soft-sphere forces contribute more towards heat current than the harmonic ones. Any explanation of this observation in terms of normal modal energy is yet to emerge.

\section{Conclusions \& Discussions}
The present manuscript relaxes the assumption of point particles prevalent in the study of thermal transport characteristics in one-dimensional $\Phi^4$ chains. Having a finite dimension, the particles of the modified chain, termed as the $\Phi^{4C}$ chain, can now ``collide'' with each other. Collisions have been modeled through an anharmonic soft-sphere potential because of which two particles strongly repel each other upon coming closer than a threshold value. This makes $\Phi^{4C}$ chains a closer one-dimensional approximation to real life systems, such as CNTs and nanowires, than $\Phi^4$ chains. It must be noted that the resulting equations of motion are stiff, necessitating very small time steps for obtaining consistent solutions. Allowing collisions between the particles results in -- (i) a drastic alteration of the temperature profiles, (ii) a significant reduction in thermal conductivity at low temperatures, and (iii) an increase in thermal conductivity beyond the inversion temperature. 

Like the traditional $\Phi^4$ chain, we observe that the $\Phi^{4C}$ chains obey Fourier's law. Except for this similarity, the thermal transport characteristics of the two chains are vastly different. The first set of differences arises in the temperature profile of the two chains -- the boundary temperature jumps typically present in $\Phi^4$ chains at low temperatures are, for all practical purposes, absent in $\Phi^{4C}$ chains. Further, the deviation from linearity is smaller in the $\Phi^{4C}$ chains at these temperature ranges. These attractive properties along with the ability to better represent real-life one-dimensional systems makes $\Phi^{4C}$ chains more suitable for studying multiscale thermal transport behavior at low temperatures. 

Perhaps, the most contrasting results arise for thermal conductivity. At low temperatures, where harmonic effects dominate, thermal conductivity of both $\Phi^4$ and $\Phi^{4C}$ chains are high, with $\Phi^4$ chains having a relatively larger thermal conductivity. However, with increasing temperature, while in $\Phi^4$ chains thermal conductivity decreases continuously, thermal conductivity in $\Phi^{4C}$ chains first decreases abruptly and then keeps increasing beyond an inversion temperature. This unique trend of $\Phi^{4C}$ chains is typically absent in other well established momentum non-conserving one-dimensional chains. The reason behind these observations have a well grounded explanation in terms of the energy transported by the normal modes of vibration. Looking at the dynamics in Fourier space, we see that the interaction between the different normal modes is more in $\Phi^{4C}$ chains than in $\Phi^4$ chains because of the collisions between the particles. This results in quicker redistribution of energy from the lowest modes to the higher modes of vibration. At low temperatures, where thermal conductivity is highly dependent on the energy transported by the lowest modes of vibration, a quick redistribution of energy from the lowest modes causes $\Phi^{4C}$ chains to have a reduced thermal conductivity than $\Phi^4$ chains. 

However, this line of argument cannot explain the rise in thermal conductivity post the inversion temperature. So, what is the underlying cause behind this rise? To answer this question, we split the heat current into two parts, and look explicitly at the contributions arising from the harmonic and anharmonic inter-particle forces. As collisions tend to increase with increasing temperature, the contribution of anharmonic inter-particle forces towards the total heat current exceeds that of harmonic forces resulting in larger thermal conductivity. 

To conclude, the $\Phi^{4C}$ chains proposed in this work have the features of both momentum non-conserving systems -- such as finite thermal conductivity, satisfying Fourier's law, etc. -- and momentum conserving systems -- such as increase in thermal conductivity upon increasing the temperature -- and may, therefore, become an important tool to study thermal transport in real-life systems. This work sets the stage for studying the effects of including collisions in momentum-conserving nonlinear chains such as an FPU chain. In an FPU chain, where the heat current occurs because of both harmonic and anharmonic inter-particle forces, it will be interesting to see if the additional anharmonicity caused by soft-sphere collisions have any bearing on both modal energy recurrence and anomalous heat transport.

\section{Acknowledgment}
Support for the research provided in part by Indian Institute of Technology Kharagpur under the grant DNI is gratefully acknowledged. Authors also gratefully acknowledge Prof. Baidurya Bhattacharya of Indian Institute of Technology Kharagpur for providing insightful comments on the manuscript.

\bibliography{apssamp}

\end{document}